# The SNR W28 at TeV Energies

G.P. Rowell[1], T. Naito[2], S.A. Dazeley[3], P.G. Edwards[4], S. Gunji[5],
T. Hara[2], J. Holder[1], A. Kawachi[1], T. Kifune[1], Y. Matsubara[8],
Y. Mizumoto[9], M. Mori[1], H. Muraishi[10], Y. Muraki[8], K. Nishijima[7],
S. Ogio[6], J.R. Patterson[3], M.D. Roberts[1], T. Sako[8],
K. Sakurazawa[6], R. Susukita[11], T. Tamura[12], T. Tanimori[6],
G.J. Thornton[3], S. Yanagita[10], T. Yoshida[10] and T. Yoshikoshi[1]

[1] *Institute for Cosmic Ray Research, University of Tokyo, Tokyo 188-8502, Japan*
[2] *Fac. of Management Information, Yamanashi Gakuin University, Yamanashi 400-8575, Japan*
[3] *Dept. of Physics and Math. Physics, University of Adelaide 5005, Australia*
[4] *Institute of Space and Astronautical Science, Kanagawa 229-8510, Japan*
[5] *Dept. of Physics, Yamagata University, Yamagata 990-8560, Japan*
[6] *Dept. of Physics, Tokyo Institute of Technology, Tokyo 152-8551, Japan*
[7] *Dept. of Physics, Tokai University, Kanagawa 259-1292, Japan*
[8] *Solar-Terrestrial Environment Lab., Nagoya University, Aichi 464-8601, Japan*
[9] *National Astronomical Observatory of Japan, Tokyo 181-8588, Japan*
[10] *Faculty of Science, Ibaraki University, Ibaraki 310-8512, Japan*
[11] *Institute of Physical and Chemical Research, Saitama 351-0198, Japan*
[12] *Faculty of Engineering, Kanagawa University, Kanagawa 221-8686, Japan*

**Abstract.** The southern supernova remnant (SNR) W28 was observed in 1994 and 1995 by the CANGAROO 3.8m telescope in a search for multi-TeV gamma ray emission, using the Čerenkov imaging technique. We obtained upper limits for a variety of point-like and extended features within a $\pm1°$ region and briefly discuss these results, together with that of EGRET within the framework of a shock acceleration model of the W28 SNR.

## INTRODUCTION

W28 is a composite SNR (mixed or M-type) with centrally filled X-ray and optical emission and limb brightened or shell-like radio emission [10,6]. It lies at a distance of about 1.8 kpc (from $\Sigma$-D, although kinematic arguments place a higher figure of 4 kpc), has an age of between 3.5-15$\times 10^4$ yrs, and evolution consistent with the radiative or Sedov phases. The radio shell ($\sim 1°$ diameter) is dominated by the northern half and over 40 maser emission (1720 MHz) sites have been identified indicating strong interaction with a molecular cloud [2]. The

ROSAT X-ray emission is well explained by a thermal model, but recent ASCA data hint at non-thermal emission in the southwest region [13]. A flat spectrum (integral index −0.9) unidentified EGRET source, 3EG J1800-2338 (0.32° 95% error circle radius), [4] is centred on the southern radio edge. W28 and the EGRET source are a strong example of an EGRET source/SNR association [11]. The radio pulsar PSR J1801-23 at the northern SNR edge is not thought to be associated with W28 given the difference in distances of this and the SNR [5].

SNR are thought primarily responsible for the acceleration of galactic cosmic-rays (CR) and W28 is a good southern hemisphere example of such a site. Gamma-ray emission can be produced from one, or a combination of hadronic ($p + p \rightarrow \pi^\circ \rightarrow 2\gamma$) and electronic (inverse Compton boosting of ambient soft photons and bremsstrahlung) processes extending up to TeV energies. The emission at TeV energies (and non-thermal X-ray synchrotron emission) is therefore a tracer of CR acceleration.

## DATA ANALYSIS AND RESULTS

We used the 3.8 metre telescope of CANGAROO [3] in a search for TeV gamma-ray emission from the W28 region over two observation seasons (1994 and 1995). The imaging camera on this telescope has a field of view $\sim 3°$ on a side and we have used an analysis that maintains a roughly constant gamma-ray selection power for sources located within a $\sim 1° \times 1°$ area [12] of the telescope tracking position. ON source data were complemented by a set of OFF source data (tracking position displaced in right ascension only) for background comparison. Following removal of data under the influence of weather and instrumental effects, a total (for 1994 and 1995) of 57.5 hours ON and 53.5 hours OFF source data were accepted for analysis.

The image cuts on data are based on a combination of the Hillas image orientation, location and size parameters (see [9] for a technique summary). W28, if a TeV emitter, may contain both point-like and extended features, requiring a detailed study of the off-axis performance of the CANGAROO 3.8m camera. Simulations of the telescope/camera combination reveal a decreasing gamma-ray selection efficiency of the cuts for off-axis point sources, due to camera-edge effects. It is possible to maintain an improved gamma-ray cut efficiency over the camera using a combination of cuts that are dependent on the location of the assumed source. One of these cuts, $D$, characterises the distance between the assumed and reconstructed source position for each event:

$$D = \sqrt{\left(\frac{miss}{\sigma_{miss}}\right)^2 + \left(\frac{dis - dis_{ex}}{\sigma_{dis}}\right)^2} \qquad (1)$$

where the expected *dis* of an image, $dis_{ex} = 1.25(1.0 - \frac{width}{length})$, is dependent on the image elongation. The standard deviations are given by $\sigma_{dis} = 0.21 + 0.09d$, ie.

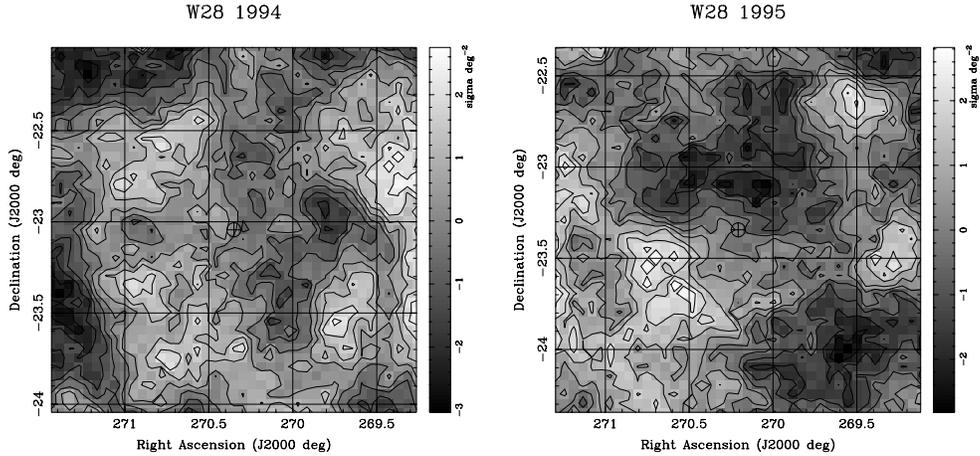

**FIGURE 1.** Skymaps of point source statistical significance (normalised) for the ON–OFF source excess over a $\pm 1°$ field after application of all cuts (table 1). The tracking positions for each year's data differs by $\sim 0.34°$.

dependent on source distance from the camera centre $d$, and $\sigma_{miss} = 0.09$ for all $d$. A *length* cut dependent upon source position is also applied. These cuts are all selected *a priori* using Monte Carlo simulations of the CANGAROO 3.8 metre telescope and camera, and provide a quality factor of $\sim 4$ at $\sim 40\%$ gamma-ray cut efficiency for any point source located within a $\pm 1°$ area. A full description of this analysis is given by [12].

A number of sites (point and extended, listed in table 1) within the W28 region were searched for TeV gamma-ray emission. The tracking position of 1994 data was PSR J1801−23 and that for 1995 was a radio position labelled 'A83', defined by [1]. Skymaps (normalised ON−OFF excess in sigma for a point source as a function of assumed source position) are presented in figure 1, and reveal no statistically significant excesses. Upper limits to the TeV gamma-ray flux at the $3\sigma$ level (listed in table 1) were calculated. For the extended source examples, the events satisfying the cuts at source positions within a radius of interest were summed for ON and OFF source data. A slightly different gamma-ray trigger efficiency was used for each year's data to account for a lower event rate in 1995 compared to 1994, and all results were normalised to a 1.5 TeV energy threshold (representing the energy at the half-maximum of the distribution of triggered energies).

## MODEL COMPARISON

We make a comparison of the EGRET results (spectrum from [7]) and our upper limit for an extended source centred on A83 from 1994 data (the lowest of our extended source upper limits) with a model of the TeV gamma-ray flux due to the decay of neutral pions [8] in figure 2. The model flux will scale according to

**TABLE 1.** Summary of the 3$\sigma$ flux upper limits from several sites/features within the W28 region.

| Feature | Flux($\geq$ 1.5 TeV) ph cm$^{-2}$ s$^{-1}$ | |
|---|---|---|
| | 1994 Data | 1995 Data |
| Radio position A83[a] | $< 3.36 \times 10^{-12}$ | $< 2.95 \times 10^{-12}$ |
| Radio position A83[b] | $< 8.75 \times 10^{-12}$ | $< 6.67 \times 10^{-12}$ |
| PSR J1801$-$23[c] | $< 3.20 \times 10^{-12}$ | $< 3.32 \times 10^{-12}$ |
| Masers (E&F)[d] | $< 4.14 \times 10^{-12}$ | $< 3.47 \times 10^{-12}$ |
| 3EGJ1800$-$2338[e] | $< 8.82 \times 10^{-12}$ | $< 1.18 \times 10^{-11}$ |

a: Point source at radio position A83, defined by [1].

b: Extended source of radius 0.35° centred on A83.

c: Point source at pulsar position [5].

d: Point source at average position of the two strongest maser sites E and F [2].

e: Highest pointlike significance within EGRET 95% error circle (0.32°) [4].

$F_\gamma \propto \frac{E_t(10^{51}\text{erg}) \, n(\text{cm}^{-3})}{d^2(\text{kpc}^2)}$, where values of the total energy available for CR production, $E_t = 0.01 \rightarrow 0.1$, the distance to the remnant, $d = 1.8 \rightarrow 4.0$ kpc, and the density of ambient matter, $n = 1.3$ cm$^{-3}$, are published ranges. The results of scaling the model flux according to these range of values are defined by the hashed region in figure 2. Results when assuming a much higher matter density of $n = 20$ cm$^{-3}$ in combination with favourable values of $E_t$ and $d$ are indicated by the thick dot-dashed line. A proton injection spectrum of $-2.1$ (differential) and cutoff energy $10^{14}$eV has been used in the model, ie. consistent with shock acceleration. When assuming a high value of $n$, the model flux is able to meet the EGRET data without violating our upper limit.

## DISCUSSION

A search for TeV gamma ray emission from the W28 region was carried out on data taken in 1994 and 1995 with the CANGAROO 3.8m telescope. No evidence for point-like or extended TeV $\gamma$-ray emission from a number of sites in the W28 region was found. We compare the lowest of our extended source upper limits to the predicted TeV gamma ray emission from $\pi^\circ$ decay. From figure 2, the EGRET flux may be accounted by $\pi^\circ$ decay gamma-rays alone, if we assume a high ambient matter density ($n \sim 20$ cm$^{-3}$). Such a density is possible for W28, given the presense of a nearby molecular cloud and maser emission. However, an accelerated electron component from bremsstrahlung and inverse Compton scattering may also contribute. Further studies of results at X-ray energies (for e.g. [13]), may hint at the level of such components. A more detailed investigation of model parameters including spectral cutoffs is underway.

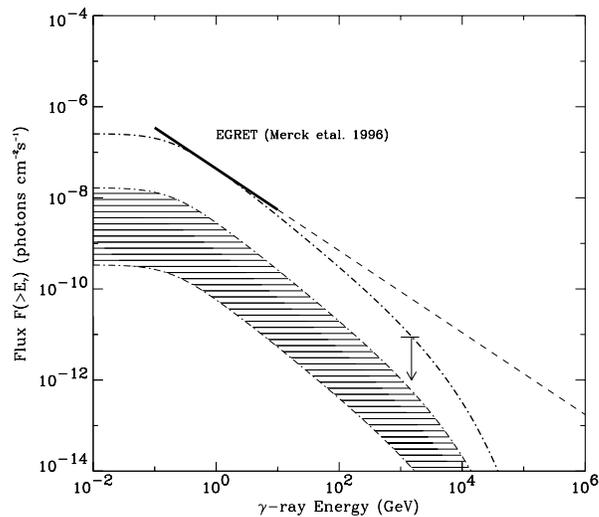

**FIGURE 2.** Comparison of our upper limit (extended source at A83 for 1994 data) and the EGRET flux of 3EG J1800−2338 [7] with a model predicting the TeV gamma-ray flux due to $\pi^\circ$ decay (hashed area and single dot-dashed line, [8]). See text for details.

## ACKNOWLEDGEMENTS

This work is supported by a Grant-in-Aid in Scientific Research from the Japanese Ministry of Science, Sports and Culture, and also by the Australian Research Council. GR acknowledges the receipt of a JSPS postdoctoral fellowship.